\documentclass[reprint,amsmath,amssymb,aps,superscriptaddress,nofootinbib]{revtex4-1} 
 
\usepackage[dvipsnames]{xcolor}

\usepackage[T1]{fontenc}
\usepackage[utf8]{inputenc}
\usepackage{graphicx}
\usepackage{dcolumn}
\usepackage{bbm}
\usepackage{float}
\usepackage{amsmath}

\usepackage[normalem]{ulem} 

\usepackage{placeins}

\newcommand{\jgs}[1]{\textcolor{red}{#1}}

\begin{document}

\title{Machine learning in phase transition analysis of  lattice quantum gravity}

\author{J.~Ambjorn}
\affiliation{{\small{The Niels Bohr Institute, Copenhagen University, Blegdamsvej 17, DK-2100 Copenhagen Ø, Denmark.}}}
\affiliation{%
{\small{IMAPP, Radboud University, Houtlaan 4, 6525 XZ Nijmegen, The Netherlands}}}

\author{Z. Drogosz}%
 \affiliation{%
{\small{Institute of Theoretical Physics, Jagiellonian University, \L ojasiewicza 11, Kraków, PL 30-348, Poland.}}}
 \author{J. Gizbert-Studnicki}%
 \email{jakub.gizbert-studnicki@uj.edu.pl}
 \author{A. Görlich}%
\affiliation{%
{\small{Institute of Theoretical Physics, Jagiellonian University, \L ojasiewicza 11, Kraków, PL 30-348, Poland.}}}%
\affiliation{%
{\small{Mark Kac Center for Complex Systems Research, Jagiellonian University, \L ojasiewicza 11, Kraków, PL 30-348, Poland.}}}

 \author{D. Németh}%

\affiliation{%
{\small{IMAPP, Radboud University, Houtlaan 4, 6525 XZ Nijmegen, The Netherlands}}}%

\author{M. Reitz}%
\affiliation{{\small{TNO – Netherlands Organization for Applied Scientific Research, High Tech Campus 21, 5656 AE Eindhoven, The Netherlands.}}}

\date{\today}%

\begin{abstract}
Using numerical data coming from Monte Carlo simulations of four-dimensional Causal Dynamical Triangulations, we study how automated machine learning algorithms can be used to recognize transitions between different phases of quantum geometries observed in lattice quantum gravity. We tested seven supervised and seven unsupervised machine learning models and found that most of them were very successful in that task, even outperforming standard methods based on order parameters. 
\end{abstract}

\maketitle

\section{Introduction}
Phase structure recognition and phase transition analysis constitute an important problem in many lattice quantum field theories. 
In particular in the context of quantum gravity, the prospective  continuum limit, ideally
one consistent with the putative UV fixed point of quantum gravity postulated by the asymptotic safety conjecture \cite{weinberg} and sought in the functional renormalization group approaches \cite{reviewFRG1,reviewFRG2,reviewFRG3}, should be related to a phase transition point of lattice theory \cite{RGflow,RenormalCDT,RGflowNew}. At the same time, one aims to reproduce the correct infrared limit, consistent with a small quantum perturbation of
general relativity (GR). Therefore, recognizing different phases of quantum geometry and analyzing phase transitions using Monte Carlo (MC)
data remains a vital task. For example, in the Causal Dynamical Triangulations (CDT) approach, one observes a rich phase structure, with four different phases of quantum geometry, of which only three were initially recognized \cite{RecUniv,CDTHL}, and it took more than a decade before the fourth one (the so-called phase $C_b$ or the ``bifurcation phase'') was discovered \cite{Signature}. This was due to the too narrow set of order parameters used at that time, which were insensitive to the bifurcation phase transition. 
This highlights a limitation of conventional approaches, which are suitable when the relevant order parameter is known.
Machine learning (ML) techniques answer the question whether the collected samples in a high-dimensional data set lie on distinguishable manifolds. They may learn nonlinear boundaries that are difficult to formulate manually, detect distributional changes that hint at new phases, and provide candidate order parameters.
It is thus tempting to ask whether ML techniques can be used to give some insight into the nature of the observed phase transitions and (prospectively) to automatically explore the CDT parameter space in search of potential new phases.

The pioneering works showed the feasibility of phase identification in condensed-matter systems through supervised learning directly from Monte Carlo configurations \cite{Carrasquilla:2016oun} and through unsupervised learning on features extracted by Principal Component Analysis \cite{wang2016discovering}. It was shown that neural networks may identify unexplored phase transitions based on data with unknown or incorrect labeling
\cite{vanNieuwenburg:2016zsd}.
Soon afterwards, both supervised \cite{Wetzel:2017ooo} and unsupervised learning \cite{wetzel2017unsupervised} were used for phase transition identification in lattice field theory.
However, to the best of our knowledge, machine learning has not been used so far in the context of lattice quantum gravity, where phase transitions are somewhat atypical, sharing certain features of both first- and higher-order transitions, or even exhibiting characteristics of transitions of a topological nature \cite{TopologyInduced}. It is therefore interesting to investigate whether these methods can also be successfully applied in this context. 

In this work, we take the first step in this direction by using ML methods to analyze phase transitions already observed in CDT.
To verify the robustness of the approach and determine whether different algorithms yield consistent transition signals, we tested seven supervised and seven unsupervised models. This also allowed us to examine whether simple and highly interpretable models, such as, e.g., {\it Logistic Regression}, can perform as well as more sophisticated but less interpretable approaches, such as, e.g., {\it Neural Networks}.
Finally, we investigated how the choice of features measured in Monte Carlo simulations and subsequently used in ML affects the results, and whether it provides clues for identifying appropriate order parameters for each of the analyzed phase transitions.

\section{Causal Dynamical Triangulations}
Causal Dynamical Triangulations (CDT) is an approach to quantizing gravity using a nonperturbative lattice quantum field theory framework; see \cite{DynTriangLQG} for its detailed formulation and \cite{oldReview,RenateReview, torusReview} for reviews. CDT regularizes the (formal) quantum-gravitational path integral over geometries  by a sum over a countable set of triangulations constructed in four dimensions from two types of elementary simplicial building blocks with fixed edge lengths. An important assumption of CDT is the restriction of the path integral to globally hyperbolic geometries (triangulations), which can be foliated into spatial hypersurfaces of identical fixed topology.
The triangulations are combinatorial, i.e., each simplex is uniquely specified by its set of vertices, and simplices are glued along entire faces only. This ensures that the resulting complex is a genuine simplicial manifold of fixed global topology, without self-identifications or conical singularities beyond the standard Regge-type curvature defects.

In each triangulation, a spatial slice with integer lattice time coordinate $t$ is constructed by gluing together equilateral tetrahedra so
that the chosen spatial topology is preserved. The neighboring spatial slices at $t$ and $t + 1$ are connected by timelike edges that together with the tetrahedra form four-dimensional simplices with either four vertices on one time-slice and one vertex on the other one – the (4,1)-simplex – or three vertices on one time-slice and two on the other one – the (3,2)-simplex. The four-simplices are internally flat. Curvature is defined by deficit angles around two-dimensional ``bones'' (triangles), so by gluing simplices together along their three-dimensional faces nontrivial geometries emerge.

The triangulations are summed over with a weight dependent on the Einstein-Hilbert action, which for a piecewise flat triangulation becomes the Regge action \cite{regge}. In CDT, the Regge action takes the simple form of a linear combination of certain global numbers characterizing a triangulation, i.e., the total number of vertices, denoted $N_0$, and the total numbers of the two types of four-simplices, denoted $N_{41}$ and $N_{32}$, are weighted by three dimensionless coupling constants,  $\kappa_0$,  $\kappa_4$ and $\Delta$, related to the gravitational constant, the cosmological constant and the asymmetry of lengths of spacelike and timelike links in the lattice, respectively.
The CDT time foliation  enables a well-defined Wick rotation from  Lorentzian to Euclidean metric signature, after which CDT becomes a statistical theory of triangulated random geometries, which can be investigated using MC
techniques. In a MC simulation, we let the total number of $(4,1)-$simplices fluctuate around a fixed target volume $\bar N_{41}$
(the average number of $(4,1)$-simplices)
by fine-tuning the lattice cosmological constant $\kappa_4\to \kappa_4^c(\kappa_0,\Delta,\bar N_{41})$. This results in a two-dimensional parameter space: $(\kappa_0,\Delta)$. 

\begin{figure}
\centering
\includegraphics[width=0.45\textwidth]{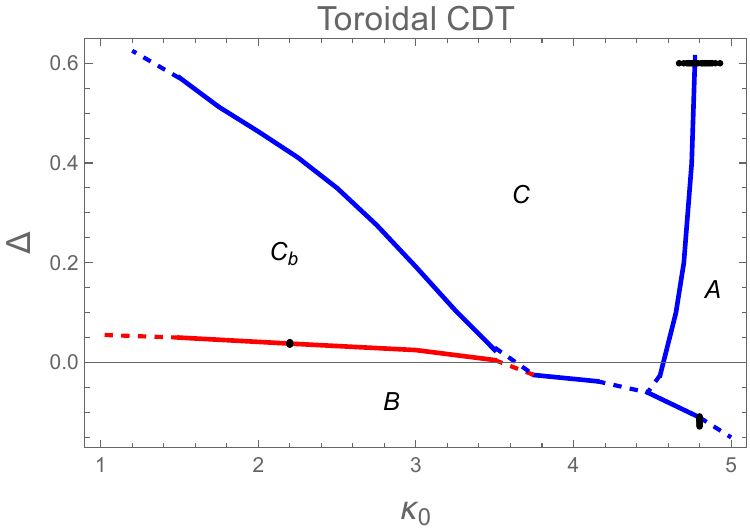}
\caption{The phase diagram of the toroidal CDT. Solid lines denote measured phase transition lines, where first-order transitions are shown in blue while higher-order transitions in red. Dashed lines are extrapolations. Black dots indicate the points where actual Monte Carlo simulations were performed for $\bar N_{41}=100\mathrm{k}$ and used in ML methods discussed in this work. }
\label{fig_Phases}
\end{figure}

For two choices of the fixed spatial topology, that of a three-sphere and a three-torus, the parameter space has been thoroughly scanned in search of distinct phases of dynamically emerging quantum geometry. Four phases (named $A$, $B$, $C$ and $C_b$) were found, and they are separated by both first- \cite{SamoLong,PhasestructureTorus,CriticalPhenomena,TopologyInduced} and higher-order \cite{SamoPRL,SamoLong,ExploringCoumbe,Czelusta} phase transition lines; see Fig.~\ref{fig_Phases}. In the following, we will focus on the toroidal CDT case, and we will analyze three of the phase transitions, namely the $A-B$, $A-C$ and $B-C_b$, which, using standard order parameter approach, were classified as a first-order \cite{TopologyInduced}, a ``weak'' first-order \cite{CriticalPhenomena}, and a higher-order transition \cite{Czelusta}, respectively.

\vspace{4pt}

\section{Numerical setup}
As the first step, Monte Carlo simulations were performed at various points of the CDT phase diagram to generate data characterizing generic quantum spacetime geometries appearing in each of the four phases (compare Fig.~\ref{fig_Phases}). 
All the simulations were performed for the fixed spatial toroidal topology and the fixed number of spatial slices $N_t=4$ (using time-periodic boundary conditions). In order to test possible volume dependence, all measurements were repeated for a set of lattice volumes ranging from $\bar N_{41}= 20\,000$ to $\bar N_{41}=600\,000$.
The measurement data obtained were subsequently used as input {\it features} for all tested machine learning algorithms. 
Since numerical CDT simulations provide a very large number of observables characterizing quantum spacetime geometries (a typical MC configuration contains hundreds of thousands of degrees of freedom), for practical reasons we chose a set of 30 features characterizing purely geometric properties of the MC-generated triangulations. The features did not include any information about the values of the CDT coupling constants or other MC simulation parameters; see  Appendix 1 for details.
In order to apply supervised machine learning algorithms and validate the results (both in supervised and unsupervised learning approaches), a ``manual'' classification of the measurement datasets into individual CDT phases was performed.
We tested three CDT phase transitions (the $A-B$, $A-C$, and $B-C_b$ transitions), whose orders had previously been determined using standard methods of statistical physics based on order parameters. For each phase transition studied, we fixed one of the CDT coupling constants in the MC simulations ($\Delta=0.6$ for the $A-C$ transition,  $\kappa_0=4.8$ for the $A-B$ transition and $\kappa_0=2.2$ for the $B-C_b$ transition), and the phase transition was triggered by changing  the other coupling constant ($\kappa_0$ or $\Delta$, respectively).
\vspace{4pt}

\begin{figure}
\centering
\includegraphics[width=\linewidth]{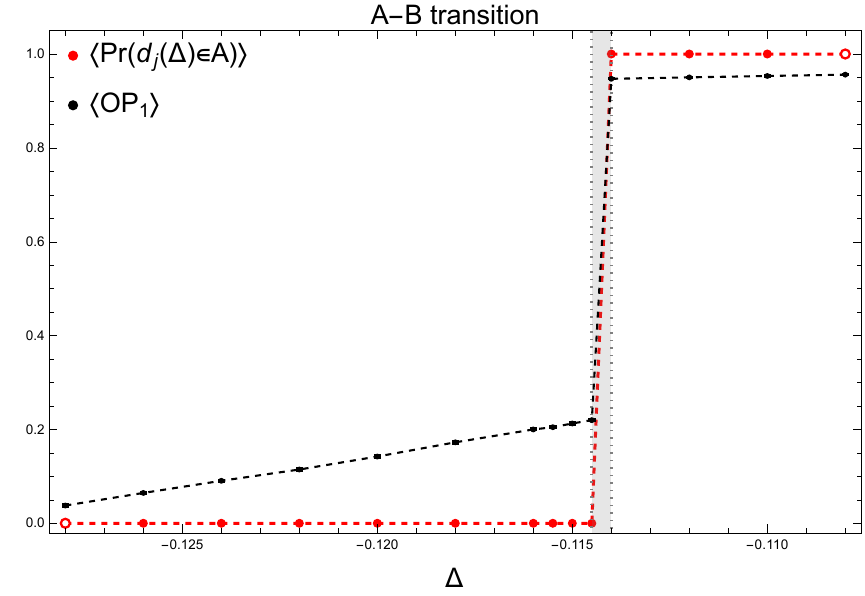}
\includegraphics[width=\linewidth]{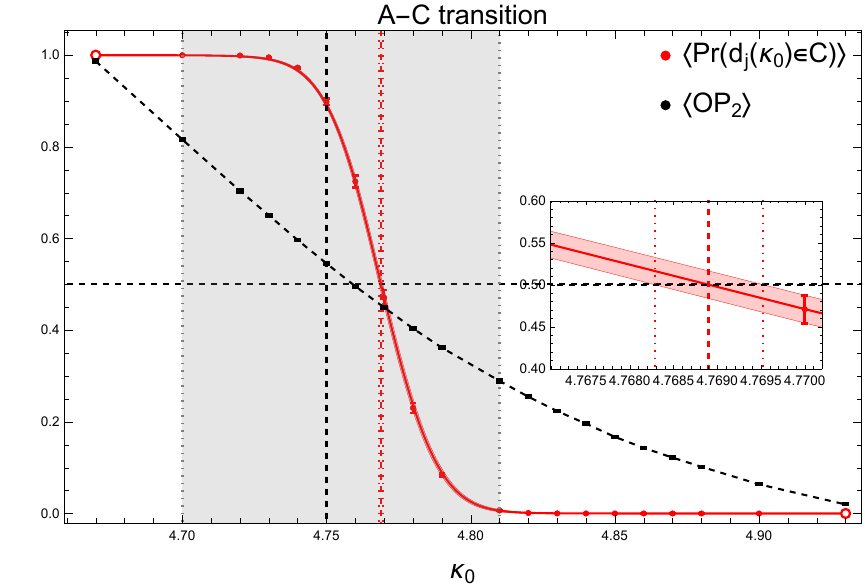}
\includegraphics[width=\linewidth]{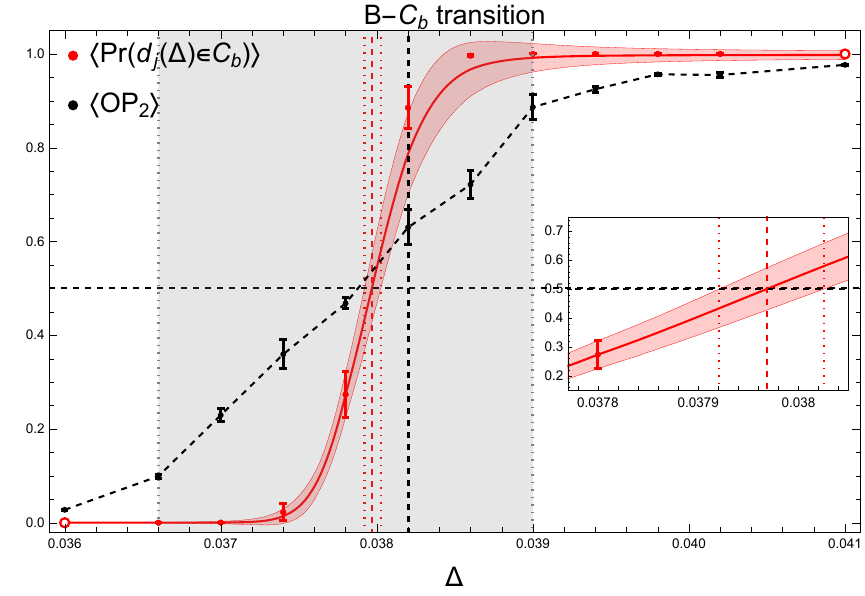}
\caption{ML analysis of the CDT phase transitions for  $\bar N_{41}=100\mathrm{k}$ using mean probabilities (red data) computed with the {\it Logistic Regression} model.
The solid red lines represent  mean probabilities obtained using histogram reweighting  applied to MC data measured closest to the transition point ($\kappa_0=4.77$ for the $A - C$ transition and $\Delta=0.0378$ for the $B-C_b$ transition). For the $A-B$ transition the histogram reweighting  does not work. Red bands denote  $1\sigma$ confidence intervals. The phase transition points estimated from the crossing of the mean probabilities with 0.5, together with their $1\sigma$ confidence intervals, are indicated by red vertical lines. For comparison, we also show the (rescaled and shifted) mean values of the standard CDT order parameters $OP_1=N_0/N_{41}$ 
and $OP_2 = N_{32}/N_{41}$ (black data).  
The phase transition points determined from  peaks of the order parameter susceptibilities  (see Fig.~\ref{fig_prob_Var_AB}) are indicated by black vertical lines, and their $1\sigma$ confidence intervals are shaded in gray. } 
\label{fig_prob_AB}
\end{figure}
\begin{figure}
\centering
\includegraphics[width=\linewidth]{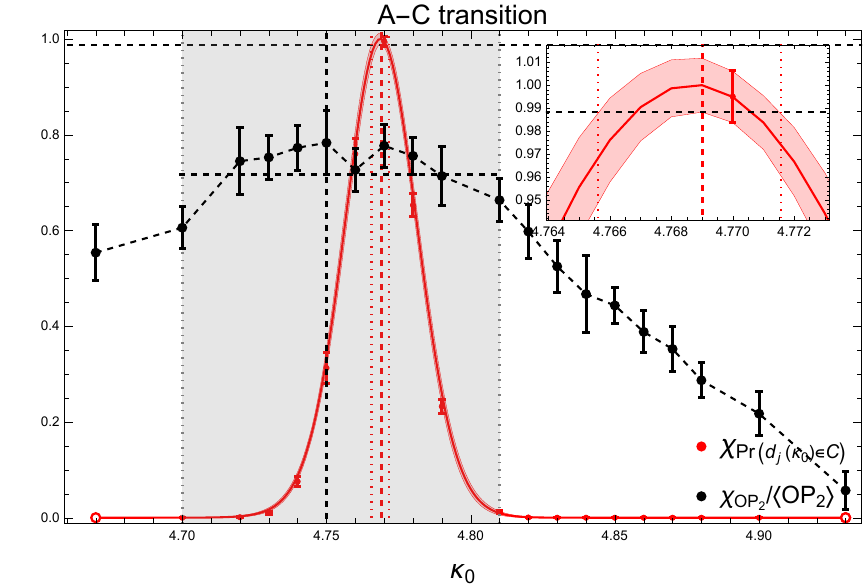}
\includegraphics[width=\linewidth]{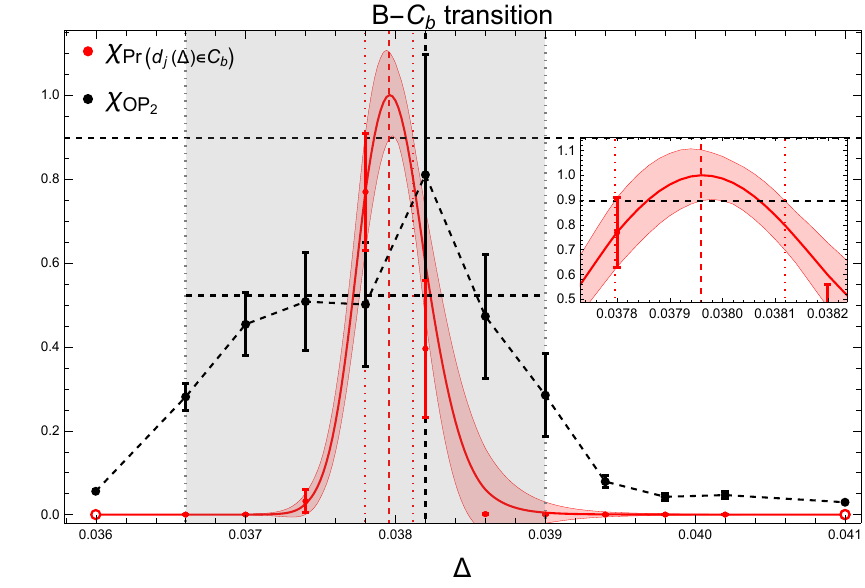}
\caption{ML analysis of the $A-C$ and $B-C_b$  phase transitions for  $\bar N_{41}=100\mathrm{k}$ using susceptibilities of probabilities (red data) computed with the {\it Logistic Regression} model. 
The solid red lines represent  the susceptibilities  obtained using histogram reweighting  applied to MC data measured closest to the transition point ($\kappa_0=4.77$ for the $A - C$ transition and $\Delta=0.0378$ for the $B-C_b$ transition). Red bands denote  $1\sigma$ confidence intervals. The phase transition points estimated from position of the peaks in the susceptibilities, together with their $1\sigma$ confidence intervals, are indicated by red vertical lines. For comparison, we also show the susceptibilities of the standard CDT order parameter $OP_2 = N_{32}/N_{41}$  (black data). 
The phase transition points determined from  peaks of the order parameter susceptibilities  are indicated by black vertical lines, and their $1\sigma$ confidence intervals are shown as gray shaded regions.  Both susceptibilities of the probabilities and the order parameters were rescaled to fit in range $[0,1]$.}
\label{fig_prob_Var_AB}
\end{figure}

For the purposes of data analysis and machine learning, we used built-in functions of \textit{Wolfram Mathematica 14.1}:
\texttt{Classify} and \texttt{ClusterClassify} for the supervised and unsupervised ML methods, respectively \cite{WolframClassify,WolframClusterclassify}.
Seven supervised ML methods ({\it Decision Tree, Gradient Boosted Trees, Logistic Regression, Nearest Neighbors, Neural Network, Random Forest, Support Vector Machines}) and seven unsupervised ML methods ({\it Agglomerate, DBSCAN, Gaussian Mixture, K-Means, MeanShift, Neighborhood Contraction, Spectral}) were tested \cite{WolframMethods}. We started by applying \texttt{``Automatic''} options, but certain machine learning algorithms required manual optimization of their hyperparameters.
To test the effectiveness of those methods in identifying the phase transitions of the CDT model, the following procedure was adopted for each phase transition studied and each MC simulation volume $\bar N_{41}$ separately:
\begin{itemize}
\item As input data for each ML algorithm, measurement results from a single MC simulation of the CDT model were selected for parameter values ($\kappa_0$ or $\Delta$) located deepest within each of the relevant phases (for instance, for the $A-C$ transition, the input data corresponded to two points, one deepest inside phase $A$ and the other one deepest inside phase $C$, respectively).
\item The ML model was then trained on a subset of the above-defined data ({\it training dataset}). 
\item The next step was to verify whether the trained ML model was capable of correctly classifying the data as belonging to the appropriate phases (for supervised learning models) or as belonging to two distinct phases (for unsupervised learning models). Testing was carried out both on the {\it training dataset} and on a larger {\it validation dataset} (not used during the learning process).
\item If the algorithm successfully passed these tests, i.e., it was able to classify or cluster the data measured deepest within the chosen phases with high accuracy ($>99.9 \%$), the trained model was subsequently applied to classify or cluster data obtained from single MC simulations located closer to the respective phase transition. For each such dataset, the mean probability (as determined by the machine learning model) of belonging to a given phase/cluster was computed, along with the variance (susceptibility) of this probability.
\item The phase transition point was then identified using two criteria: (1) the location  where  the mean probability equals $0.5$, and (2) by the presence of a peak in the  susceptibility of the classification/clustering probability.
\item The location of the phase transition predicted by the ML model was then compared with the position of the transition determined by standard techniques based on CDT order parameters, where only method (2), i.e., the peak of the susceptibility of an order parameter,  could be used.\footnote{CDT order parameters are not normalized to any fixed range; consequently, the precise location of the phase transition cannot be determined by comparing the order parameter to a fixed threshold value.}
\end{itemize}

As an example, take the $A-C$ phase transition  with fixed $\Delta=0.6$ and lattice volume $\bar N_{41}=100\,000$. Using MC simulations we prepared ML datasets for 20 values of $\kappa_0$, ranging from  $\kappa_0^{\mathrm{min}}= 4.67$ (inside phase $C$) to \mbox{$\kappa_0^{\mathrm{max}}= 4.93$} (inside phase $A$).  Then, for each of the ML models separately, the following procedure was applied:
\begin{enumerate}
\item Take all data measured for the highest value of $\kappa_0$ ($\kappa_0^{\mathrm{max}}= 4.93$), i.e., deepest inside phase $A$.
\item Take all data measured for the lowest value of $\kappa_0$  ($\kappa_0^{\mathrm{min}}= 4.67$), i.e., deepest inside phase $C$.
\item Split data from points 1 and 2 into {\it  training} and {\it validation} datasets.
\item Train the selected ML model using the training sets from points 1 and 2, 
this step also includes optimizing hyperparameters of the ML model, if necessary.
\item Check accuracy of the trained model using validation sets from points 1 and 2.
\item If the accuracy test was passed then use the trained ML model to  classify / cluster  other datasets measured for $\kappa_0$ between $\kappa_0^{\mathrm{min}}$ and $\kappa_0^{\mathrm{max}}$, i.e., closer to the phase transition point than data from points 1 and 2.
\item For each data point $d_j(\kappa_0), \  j=1,...,$ dataset length, that comes from a dataset measured for a given value of $\kappa_0$ compute the probability $\mathrm{Pr}(d_j(\kappa_0)\in C)$ that the data belong to phase $C$.
\item Compute the mean value of the probability $\langle \mathrm{Pr}(d_j(\kappa_0)\in C) \rangle$ measured for each dataset ($\kappa_0$)  and its susceptibility (variance) $\chi_{ \mathrm{Pr}(d_j(\kappa_0)\in C)}$. This step also includes error bar calculations, for which we applied a single-elimination (binned) jackknife procedure, after blocking the data to account for auto-correlation errors and choosing the block size for which the error is maximized; see \cite{ExploringCoumbe} for details.
\item Find the transition point predicted by a given ML model ${\kappa_0}_{ML}^{crit}$  using:  (1)  $\langle \mathrm{Pr}(d_j({\kappa_0})\in C)\rangle =0.5 $; see Fig.~\ref{fig_prob_AB}; and (2) the presence of a peak in the susceptibility  $\chi_{ \mathrm{Pr}(d_j(\kappa_0)\in C)}$; see Fig.~\ref{fig_prob_Var_AB}. In each case,  also compute the $1\sigma$, $2\sigma$ and $3\sigma$ confidence intervals around the estimated transition point. 
\item Compare ${\kappa_0}_{ML}^{crit}$ with ${\kappa_0}^{crit}$ measured using the peak of susceptibility of the standard CDT order parameter $OP_2$;  see Fig.~\ref{fig_prob_Var_AB}.
\end{enumerate}

Following previous works \cite{Bachtis:2020dmf,Bachtis:2020ajb}, we used Monte Carlo
histogram reweighting
\cite{Ferrenberg:1988yz, Ferrenberg:1989ui}
to improve the precision of ML phase transition identification.
Interestingly, although the method does not work for our standard observables, including the measured order parameters, it appears to perform remarkably well in most cases (with the exception of the very sharp $A-B$ transition) for the mean values and susceptibilities of the transition probabilities computed for all ML models analyzed in this work. For more details, see Appendix 2.
\vspace{4pt}

\FloatBarrier
\section{Results}

\begin{figure}[t]
\centering
\includegraphics[width=0.45\textwidth]{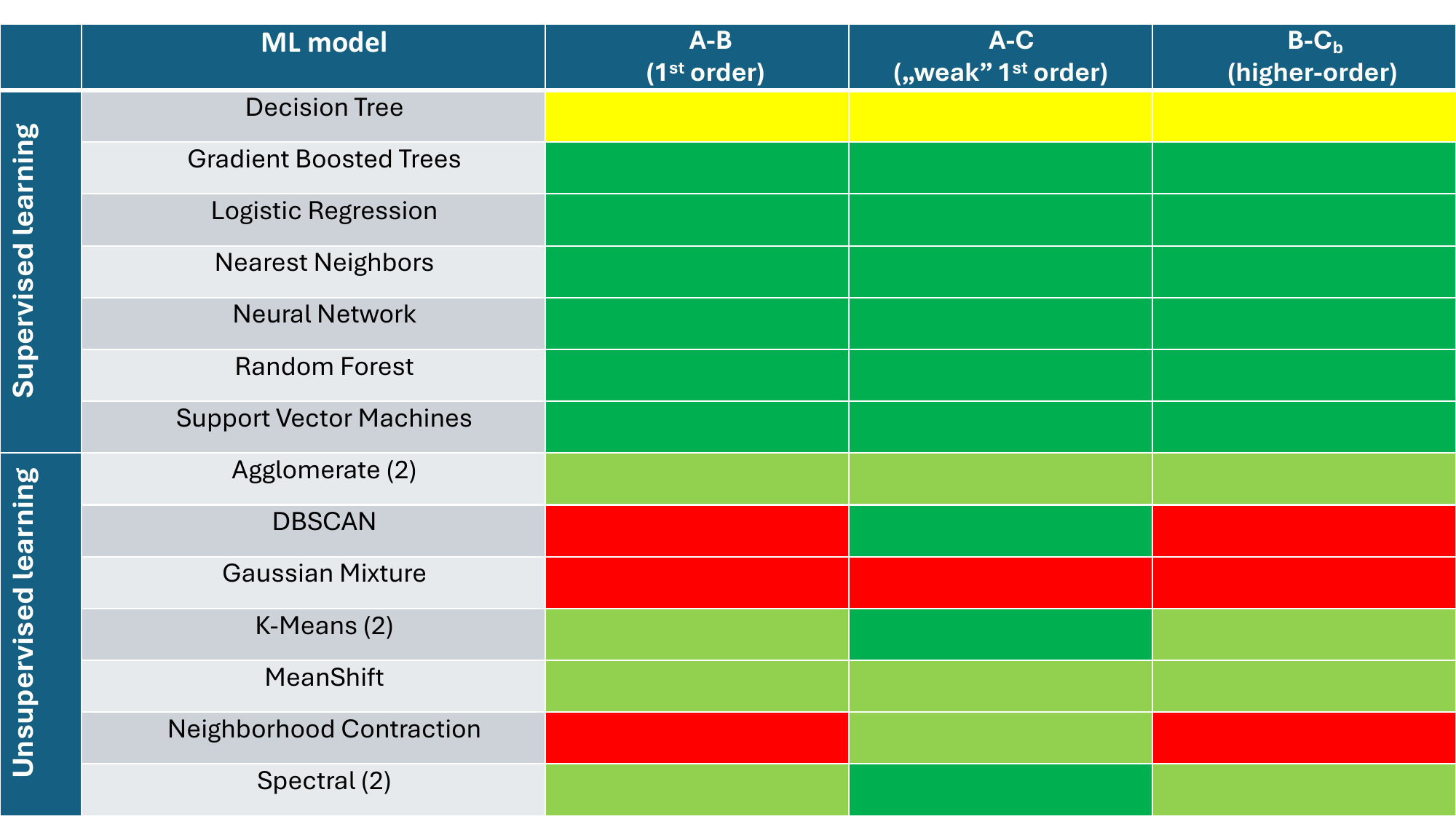}
\caption{Summary of the results obtained by ML models in the study of individual phase transitions. Legend: dark green – the model correctly identifies phase transitions without the need for ``manual'' hyperparameter optimization; light green – the model correctly identifies phase transitions but requires ``manual'' hyperparameter optimization; yellow – the model identifies phase transitions, but produces results different from those of other models and standard methods; red – the model fails to work correctly.}
\label{fig_Results}
\end{figure}

\begin{figure}
\centering
\includegraphics[width=0.45\textwidth]{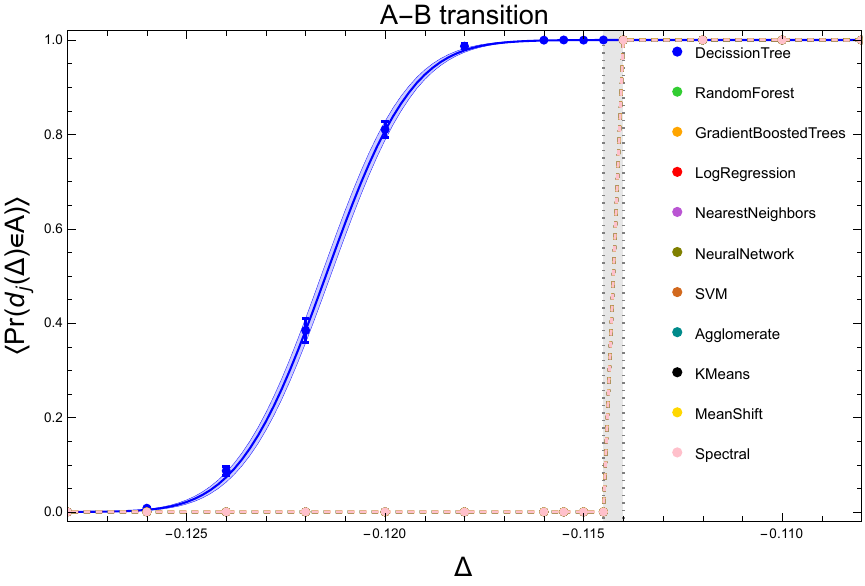}
\includegraphics[width=0.45\textwidth]{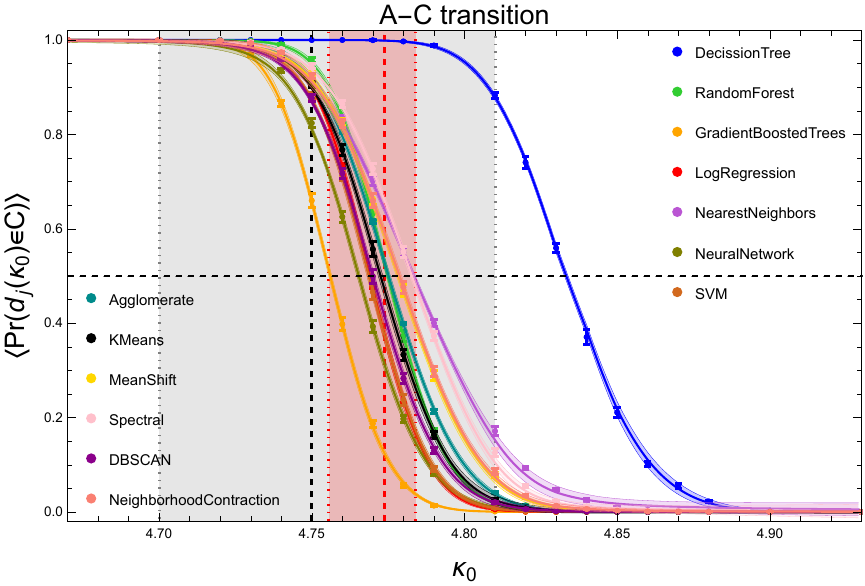}
\includegraphics[width=0.45\textwidth]{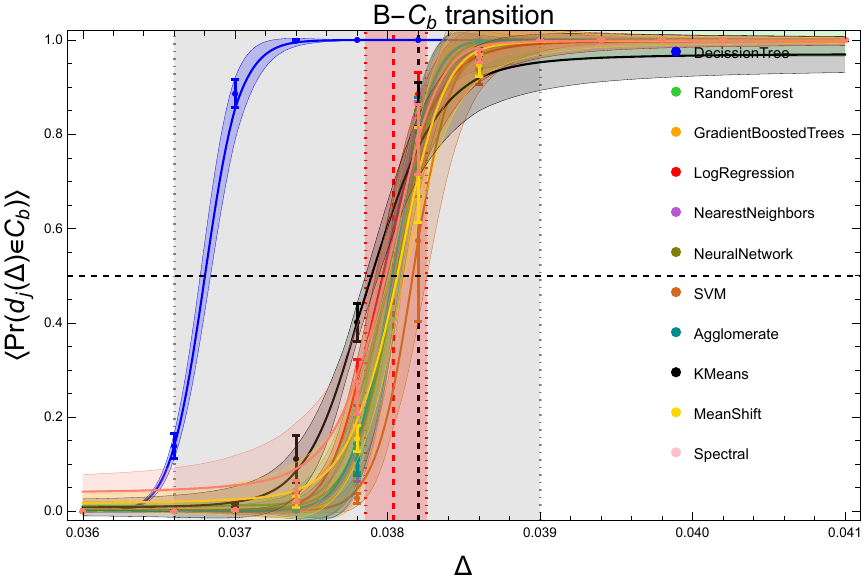}
\caption{ML analysis of the CDT phase transitions for  $\bar N_{41}=100\mathrm{k}$ using mean probabilities computed with several ML models (denoted by different colors). 
Red dashed vertical lines indicate the median positions of the phase transition points estimated across all ML models, excluding the {\it Decision Tree} model, which is treated as an outlier. These estimates are obtained from the crossing of the mean probabilities with 0.5. The red shaded region denotes the overall range of the $1\sigma$ confidence intervals (see Fig.~\ref{fig_Results_Pr_Pvar}, top panels).
For comparison, we also show the phase transition points determined from the peaks of the order parameter susceptibilities  (see Fig.~\ref{fig_prob_Var_AB}), indicated by black vertical lines, together with their $1\sigma$ confidence intervals shown as gray shaded regions. 
In the case of the $A - B$ transition, the results obtained with all models other than the {\it Decision Tree} are visually indistinguishable.}
\label{fig_PhaseAB}
\end{figure}

The performance of all tested ML models using datasets comprising  all 30  features measured in the CDT Monte Carlo simulations is summarized in Fig.~\ref{fig_Results}. For each studied phase transition, all seven supervised learning models were able to learn the classification of the individual phases of quantum gravity in CDT with high accuracy and, notably, without the need for manual hyperparameter optimization.\footnote{It is not entirely clear how much data preprocessing and hyperparameter optimization is automatically done by \textit{Wolfram Mathematica 12} build-in ML functions that we used. We could not find such information in  Wolfram's documentation.} Among them, six models ({\it Gradient Boosted Trees, Logistic Regression, Nearest Neighbors, Neural Network, Random Forest, Support Vector Machines}) produced consistent phase transition signals, in agreement with standard methods based on CDT order parameters. 
By contrast, the  {\it Decision Tree} model indicated phase transition points at different locations; see Fig.~\ref{fig_PhaseAB}. This is most likely due to the fact that in the learning process of this model perfect classification to the respective phases could be done using just one of the measured features, but such an approach proved too simplistic in recognizing phase transition point(s) correctly.

As expected, the performance of unsupervised learning models was worse. Most such models required manual hyperparameter optimization, 
with the choice depending on the type of phase transition under study. Moreover, their effectiveness depended strongly on the type of transition. Models that allowed the maximum number of clusters to be explicitly set to two (\textit{Agglomerate, K-Means, Spectral}) performed relatively well. In contrast, models lacking such an option either required manual hyperparameter tuning (\textit{MeanShift}) or failed to operate properly for certain phase transitions ({\it DBSCAN, Gaussian Mixture, Neighborhood Contraction}). In many cases (e.g., for some MC simulation volumes), these models identified too many clusters, or the resulting split of the data into clusters did not correspond to the actual division into CDT phases.
\vspace{4pt}

\begin{figure*}

\begin{minipage}{0.49\textwidth}
\includegraphics[width=\textwidth]{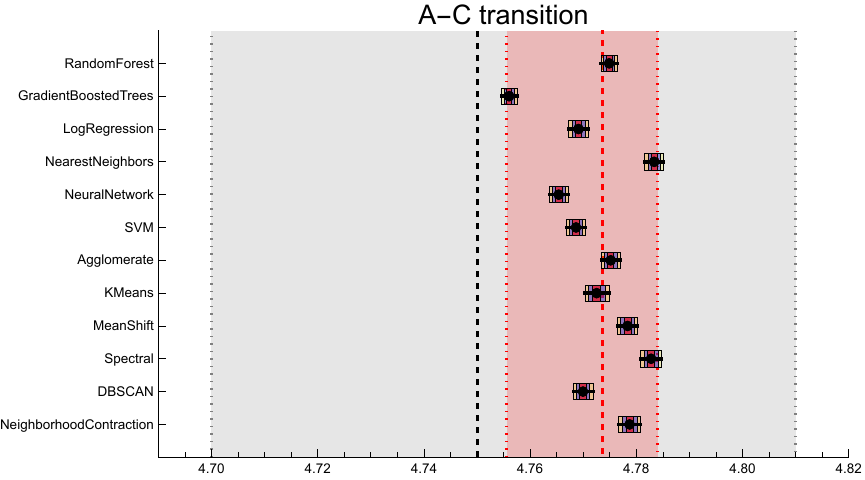}
\end{minipage}
\hfill
\begin{minipage}{0.49\textwidth}
\includegraphics[width=\textwidth]{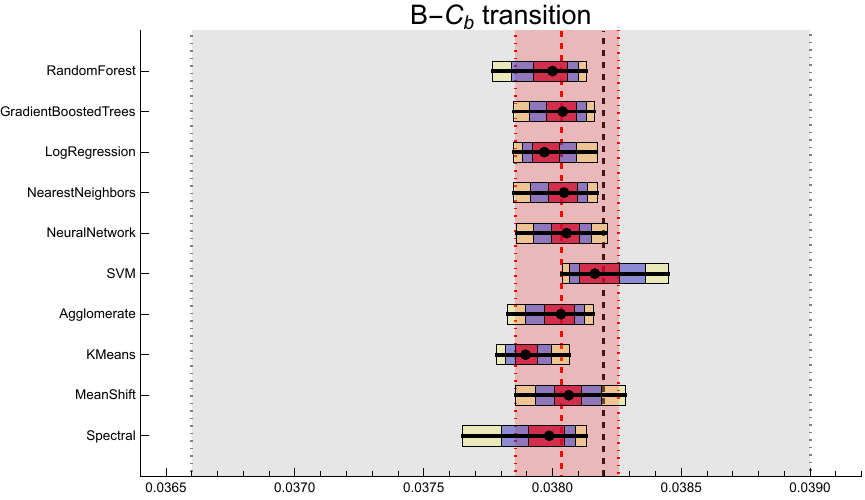}
\end{minipage}
\\[0.5cm]
\begin{minipage}{0.49\textwidth}
\includegraphics[width=\textwidth]{
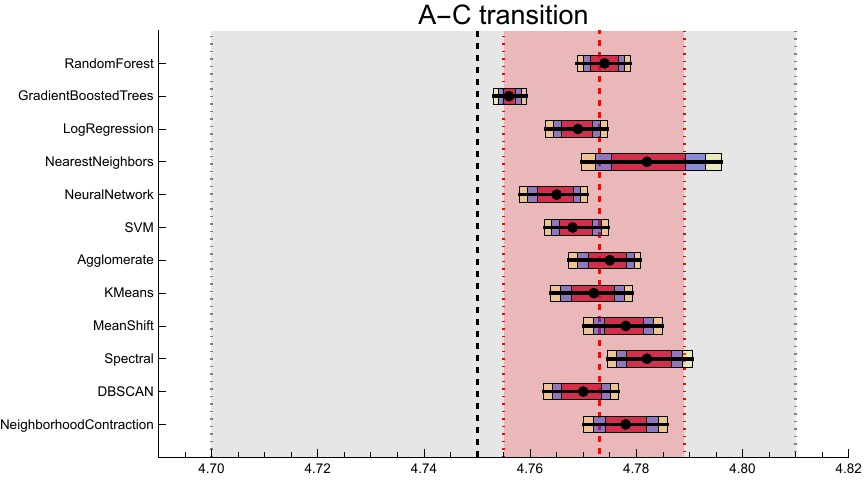}
\end{minipage}
\hfill
\begin{minipage}{0.49\textwidth}
\includegraphics[width=\textwidth]{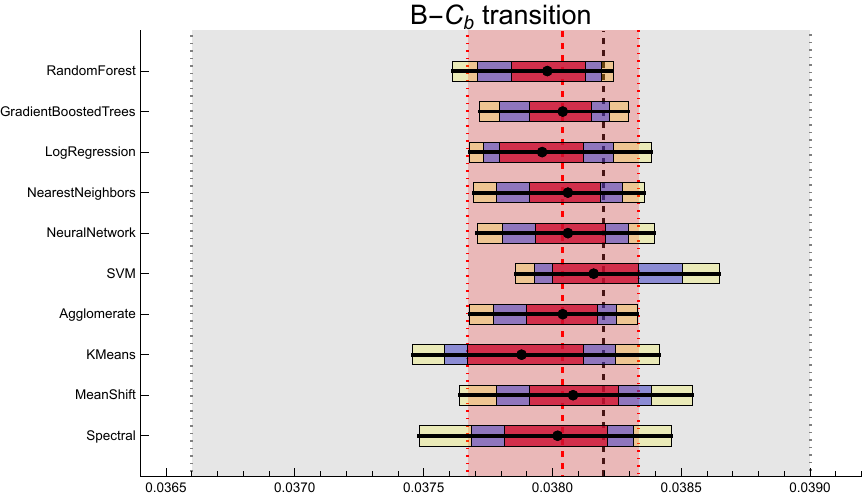}
\end{minipage}
\caption{Comparison of the results of phase transition identification using several ML models for  $\bar N_{41}=100\mathrm{k}$
based on the crossing of the mean probabilities with the value 0.5 \emph{(top panels)} or the peaks in the susceptibilities of probabilities \emph{(bottom panels)}. The $1\sigma$, $2\sigma$ and $3\sigma$ confidence intervals are  indicated by red, blue, and yellow boxes, respectively. 
Red dashed vertical lines indicate the median positions of the phase transition points estimated across all  models. The red shaded region denotes the overall range of the $1\sigma$ confidence intervals.
For comparison, we also show the phase transition points determined from  the peaks of the order parameter susceptibilities  (see Fig.~\ref{fig_prob_Var_AB}), indicated by black vertical lines, together with their $1\sigma$ confidence intervals shown as gray shaded regions.  The {\it Decision Tree} model, which is treated as an outlier, was excluded from the results.}
\label{fig_Results_Pr_Pvar}
\end{figure*}

The results of all ML models obtained for $\bar N_{41}=100\mathrm{k}$, excluding the {\it Decision Tree} model, which is treated as an outlier, are summarized in 
Fig.~\ref{fig_Results_Pr_Pvar}.
In the top panels, the phase transition point is identified using method (1), i.e., as the location where the mean classification/clustering probability equals $0.5$, while in 
the bottom panels, it is identified using method (2), i.e., from the peak of the susceptibility of the classification/clustering probability. Both methods yield consistent results.

For the $A-C$ transition, the median value obtained from all ML models using method (1) is \mbox{${\kappa_0}^{crit}_{ML(1)}=4.774^{+0.010}_{-0.018}$}, while method (2) gives \mbox{${\kappa_0}^{crit}_{ML(2)}=4.773^{+0.016}_{-0.018}$}, where the error bars are defined by the overall range of $1\sigma$ confidence intervals of the individual ML models; see 
Fig.~\ref{fig_Results_Pr_Pvar}.
These values can be directly compared with the results obtained using the standard CDT order parameters and method (2), namely $\kappa_0^{crit}=4.75^{+0.06}_{-0.05}$, which yields a more than three times wider $1\sigma$ confidence interval.

For the $B-C_b$ transition, the median values obtained using method (1) and method (2) are \mbox{${\Delta}^{crit}_{ML(1)}=0.0380^{+0.0002}_{-0.0002}$} and ${\Delta}^{crit}_{ML(2)}=0.0380^{+0.0003}_{-0.0004}$, respectively. In comparison, the standard order parameters give ${\Delta}^{crit}=0.0382^{+0.0008}_{-0.0016}$, again corresponding to a more than three times larger $1\sigma$ uncertainty.

It is therefore clearly seen that, for both transitions, the ML-based methods outperform the traditional order parameters in terms of precision by a factor of more than three for all models combined and using a very conservative method of computing ML error bars (the overall range of different models). If we compare uncertainties of the traditional method with the results of individual ML models, the difference in precision is even greater.

For the $A-B$ transition, we observe a very sharp jump in both the classification/clustering probabilities and the standard order parameters between $\Delta=-0.1155$ and $\Delta=-0.1140$; see Fig.~\ref{fig_PhaseAB}. We made every effort to increase the resolution of the MC data in the vicinity of the transition region; however, the same behavior persisted. We can therefore only conclude that the transition occurs within the above $\Delta$ interval. In this case, ML methods, although consistent with the standard approach, do not improve the precision of the phase transition measurement.



Last but not least, one can ask how the choice of features used in ML impacts the results. In the analysis presented above, we used all 30 features measured in the MC simulations. These consist of 6 {\it global} parameters characterizing the overall CDT triangulations and 24 {\it local} parameters related to the CDT time foliation within a triangulation.
The features are not independent. All global parameters can be reconstructed from the local ones, and, additionally, some of the local features are also mutually dependent; for details, see Appendix 1. One can therefore construct a set of 20 independent features. Although the choice is to some extent arbitrary, we selected a set containing only local features, and we additionally excluded $N_{14}(t)$, which satisfies $N_{14}(t)=N_{41}(t+1)$ for $t=1,2,3,4$ (assuming periodic boundary conditions).
This set of 20 independent features contains the same information as the full set of 30 features; however, in practice, the choice of features may still affect the ML results. We also tested whether restricting the analysis to only global features changes the results. In this case, we selected a set of 4 independent global features, namely $N_0$, $N_4$, $N_{41}$, and $MO$. Finally, one may also ask whether using only a single global feature is sufficient for correct phase transition identification; here we tested only $N_0$, $N_4$, and $MO$, as, by construction of the CDT volume fixing in the MC simulations, $N_{41}$ performs Gaussian fluctuations around $\bar N_{41}$ and thus has very limited impact in the phase transition studies. We restricted the above analysis to systems with  $\bar N_{41}=100\mathrm{k}$, for which we have the best resolution and statistics of the MC measurements.

The results depend to some extent on both the phase transition under consideration and the ML model used. In particular, for the $A-C$ transition, all unsupervised ML methods required either the full set of features or at least the complete set of independent features in order to work correctly. In contrast, for the $A-B$ and \mbox{$B-C_b$} transitions, no such restriction was observed, and the unsupervised ML models performed equally well when using only the global features or even a single feature (with the exception of the {\it Agglomerative} model for the $B-C_b$ transition, where the single-feature unsupervised ML analysis failed).

\begin{figure*}
\begin{minipage}{0.49\textwidth}
\includegraphics[width=\textwidth]{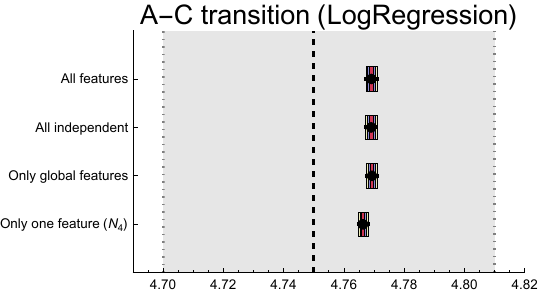}
\end{minipage}
\hfill
\begin{minipage}{0.49\textwidth}
\includegraphics[width=\textwidth]{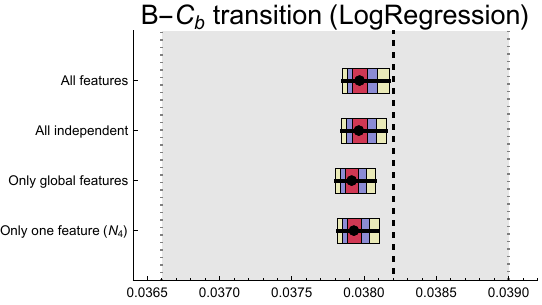}
\end{minipage}
\begin{minipage}{0.49\textwidth}
\includegraphics[width=\textwidth]{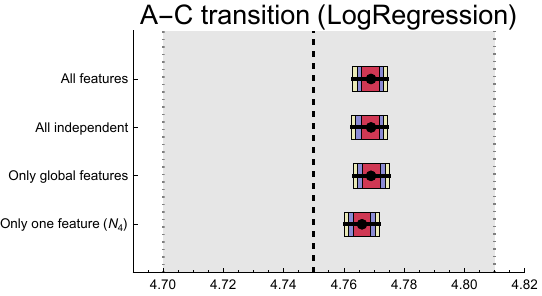}
\end{minipage}
\hfill
\begin{minipage}{0.49\textwidth}
\includegraphics[width=\textwidth]{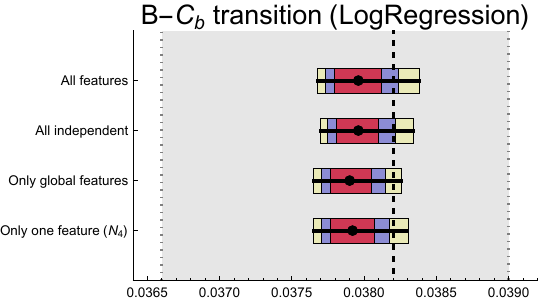}
\end{minipage}
\caption{Impact of the choice of features used in ML on the  results of the {\it Logistic Regression} model for $\bar N_{41}=100\mathrm{k}$
based on the crossing of the mean probabilities with the value 0.5 \emph{(top panels)} or the peaks in the susceptibilities of probabilities \emph{(bottom panels)}. The $1\sigma$, $2\sigma$ and $3\sigma$ confidence intervals are  indicated by red, blue, and yellow boxes, respectively.
For comparison, we also show the phase transition points determined from  the peaks of the order parameter susceptibilities  (see Fig.~\ref{fig_prob_Var_AB}), indicated by black vertical lines, together with their $1\sigma$ confidence intervals shown as gray shaded regions.}
\label{fig_Results_Pr_PVar_Choice}
\end{figure*}

All the tested supervised ML models were more flexible and worked correctly using only global features or even a single feature; 
see 
Fig.~\ref{fig_Results_Pr_PVar_Choice},
where we present the results obtained with the {\it Logistic Regression} model. The most interesting case is the $A-C$ transition, where the choice of a single feature turned out to be nontrivial. All supervised ML models worked correctly when $N_4$ was selected and failed for all other single-feature choices. We also verified that ML models using only the other global features, with $N_4$ excluded, did not work correctly.
Interestingly, this observation exactly coincides with the CDT order parameter $OP_2=N_4/N_{41}-1$ used in studies of the $A-C$ phase transition, which was shown to be the only genuinely critical parameter for this transition \cite{CriticalPhenomena}.
In the case of the $A-B$ and $B-C_b$ transitions, all ML models, including the unsupervised ones, performed well for any choice of a single feature. In some cases, the results were slightly shifted, but still remained consistent within the error bars. This is not surprising, since for these phase transitions all order parameters that can be constructed from the global features provide clear transition signals.



\section{Discussion and prospects}
We have tested seven supervised and seven unsupervised machine learning methods in the analysis of three phase transitions observed in CDT. Most of the supervised models were demonstrated to be very efficient in correctly identifying phase transition points. Some unsupervised models, especially those allowing the number of clusters to be set to two, were also very successful. Remarkably, the probabilities generated by automated ML algorithms were not only consistent with standard statistical physics methods based on order parameters, but also produced much stronger transition signals. This allowed very precise identification of the phase transition points, outperforming traditional methods by a factor of more than three within $1\sigma$ error bars.
Contrary to our expectations, more models (including the unsupervised methods) performed well in the case of the ``weak'' first-order $A-C$ phase transition than in the case of the ``typical'' first-order $A-B$ transition. The latter was correctly identified by as many models as the higher-order $B-C_b$ phase transition.

\begin{figure}
\centering
\includegraphics[width=0.45\textwidth]{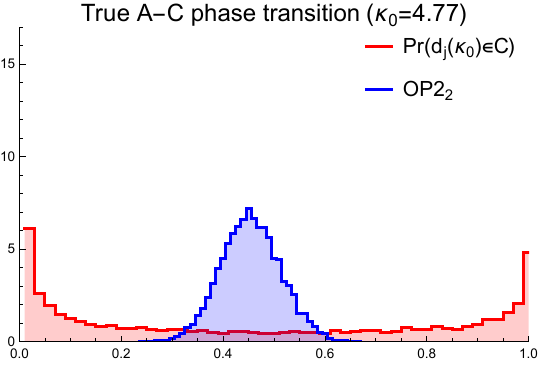}
\includegraphics[width=0.45\textwidth]{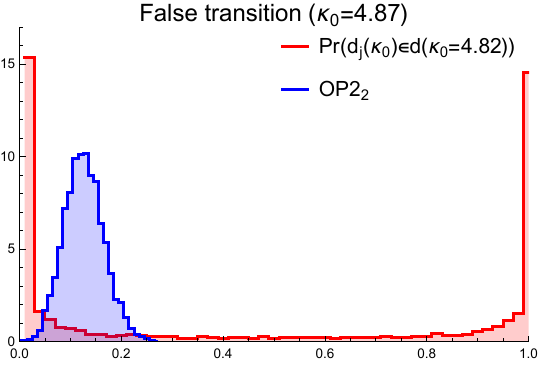}
\caption{(Probability density) histograms  of the standard CDT order parameter $OP_2$ and of the classification probabilities computed with the {\it Logistic Regression} model trained to distinguish between configurations from phases $A$ ($\kappa_0=4.93$) and $C$ ($\kappa_0=4.67$) (top), and between two different points within phase $A$ ($\kappa_0=4.93$ and $\kappa_0=4.82$) (bottom). In both cases, the histograms were measured at the $\kappa_0$ values for which the mean classification probability is close to 0.5. Results are shown for $\bar N_{41}=100\mathrm{k}$. $OP_2$ was  rescaled to fit in the range $[0,1]$.
}
\label{fig_Histograms}
\end{figure}

An interesting result that deserves further discussion is the observation that ML models perform remarkably well even when only a single feature is used. Of course, one may expect ML techniques to detect transition signals when provided with features used to construct traditional order parameters; nevertheless, the ML-based results still outperform those obtained from the standard order parameters.
On the one hand, this may help identify the correct order parameters, as illustrated by the $A-C$ transition, where the only feature that yields correct results is precisely the one consistent with the order parameter conventionally used for that transition. On the other hand, this observation must be treated with some care. In this context, the use of ML is essentially equivalent to applying a nonlinear transformation (e.g., a logistic function or the nonlinear output of a neural network) to an order parameter. Such a transformation leads to significantly stronger transition signals, but it also has important consequences for interpreting the classification/clustering probabilities as standard order parameters.
In particular, the probability distributions (histograms) of these transformed order parameters measured at, or close to, the phase transition points exhibit behavior different from that of the standard order parameters; see Fig.~\ref{fig_Histograms} (top), where we show histograms measured at the ``weak'' first-order $A-C$ transition for the transition probabilities computed with the {\it Logistic Regression} model and for the standard order parameter $OP_2$. At first sight, the behavior of the classification probabilities may appear even more characteristic of a first-order transition, since the histogram exhibits a double-peak structure, as one would normally expect, whereas no such structure is observed for $OP_2$.
However, exactly the same phenomenon is observed in the case of the higher-order $B-C_b$ transition, where double peaks should not appear in the histograms. Even more strikingly, the same behavior is also observed when the ML model is trained to distinguish between two different points within the same phase; see Fig.~\ref{fig_Histograms} (bottom), where no phase transition is present at all. This may therefore have important consequences for the analysis of susceptibilities or higher-order cumulants of the transformed order parameter, in particular for their scaling properties.
\vspace{4pt}

To summarize, the results presented here provide a promising foundation for further investigations into the applicability of ML techniques for detecting phase transitions in lattice quantum gravity approaches, 
or, more generally, in other lattice quantum field theories where phase transition signals are somewhat atypical, sharing certain features of both first- and higher-order transitions, or exhibiting characteristics of topological phase transitions.
\vspace{4pt}

Several natural directions for future research emerge:
\begin{itemize}
\item {\it Unsupervised learning approaches}. We plan to extend the analysis to larger datasets containing a greater number of measured features and revisit unsupervised learning models that failed to perform satisfactorily in this study. The collection of such data from CDT Monte Carlo simulations is currently in progress.
\item{\it Multi-phase classification}. An important extension is a test of the capability of machine learning algorithms to recognize more than just two phases simultaneously. This work is also currently in progress.
\item{\it Different spatial topologies}. Thus far, our analysis has been restricted to CDT with toroidal spatial topology. Future work will include extending the study to CDT with spherical spatial topology, which will allow us to assess the robustness of the ML methods applied.
\end{itemize}

\begin{acknowledgments}
The research was supported by a grant from the Priority Research Area DigiWorld under the Strategic Programme Excellence Initiative at Jagiellonian University. DN is supported by the VIDI programme with project number VI.Vidi.193.048, which is financed by the Dutch Research Council (NWO).
\end{acknowledgments}

\section*{Appendix 1}

Herein we list all the $30$ features which were measured in the CDT Monte Carlo simulations and then  input to the tested ML algorithms. All of the features are purely geometric observables characterizing CDT triangulations and do not include any CDT coupling constants or other parameters of the MC simulations. 
\vspace{4pt}

\paragraph{Global parameters:} 
\begin{itemize} 
\item $N_0$ – total number of vertices,
\item $N_1$ – total number of links, 
\item $N_2$ – total number of triangles, 
\item $N_4$ – total number of four-simplices, 
\item $N_{41}$ – total number of $(4, 1)$-simplices, 
\item $MO$ – maximal coordination number of vertices (maximal number of simplices sharing a vertex).
\end{itemize}

\paragraph{Local  parameters related to time-foliation:} 
\begin{itemize} 
\item  $N_{41}(t)$ – the number of $(4, 1)$-simplices with $4$ vertices in $t$ and $1$ vertex in $t+1$, 
\item $N_{14}(t)$ – the number of $(4, 1)$-simplices with $1$ vertex in $t$ and $4$ vertices in $t+1$, 
\item $N_{32}(t)$ – the number of $(3, 2)$-simplices with $3$ vertices in $t$ and $2$ vertices in $t+1$, 
\item $N_{23}(t)$ – the number of $(3, 2)$-simplices with $2$ vertices in $t$ and $3$ vertices in $t+1$, 
\item $N_0(t)$ – the number of vertices with  time coordinate $t$, 
\item $MO(t)$ – the maximal coordination number of all vertices with time coordinate $t$.
\end{itemize}
In all cases $t = 1,2,3,4$ (with periodic boundary conditions). 
\vspace{4pt}

In order  to encode the time shift symmetry of CDT, we quadrupled the dataset size by performing a time shift of all {\it local} parameters by (periodically)  changing their time  coordinates $t = (1,2,3,4) \to (4,1,2,3) \to (3,4,1,2) \to  (2,3,4,1) $. The values of the {\it global} parameters were kept unchanged.

The abovementioned observables also enable one to compute standard CDT order parameters used in  phase transition studies, i.e., $OP_1 = N_0/N_{41}$ and  \mbox{$OP_2 =N_4/N_{41}-1$}.

One should note that the 30 features described above are not independent but are constrained by the 10 relations listed below, 9 of which are linear and, in principle, can be reconstructed via Principal Component Analysis from the kernel of the feature covariance matrix. One can therefore select a set of 20 independent features. Although the choice is to some extent arbitrary, it is natural to restrict the analysis to the local features, since all global parameters can be reconstructed from them. In addition, one may eliminate four local features, e.g., $N_{14}(t)$, using the last relation given below:
\begin{align*}
& N_{0}=\sum_{t=1}^{4}N_{0}(t) , \\
& N_{41}=\sum_{t=1}^{4}\left(N_{41}(t)+N_{14}(t)\right) ,\\
& N_4=\sum_{t=1}^{4}\left(N_{41}(t)+N_{14}(t)+N_{32}(t)+N_{23}(t)\right) , \\
& N_1=3 N_0 + \frac{1}{2}N_4 ,\\
& N_2=2N_0+2N_4 , \\
& MO=\mathrm{max}\left(MO(1),...,MO(4)\right) , \\
& N_{14}(t)=N_{41}(t+1) ;\ t=1,2,3,4 \text{ (with periodic b.c.)}.  
\end{align*}

\begin{figure}
\centering
\includegraphics[width=0.45\textwidth]{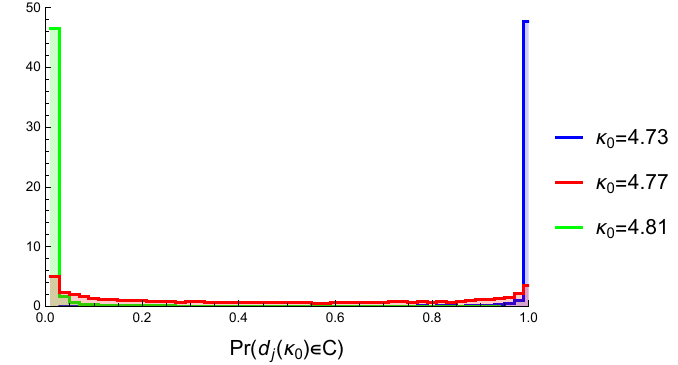}
\includegraphics[width=0.45\textwidth]{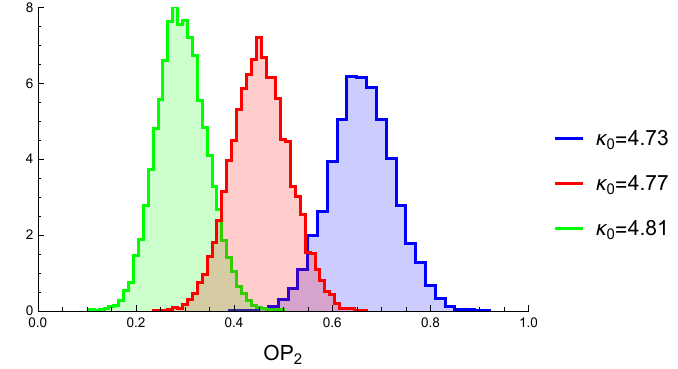}
\caption{(Probability density) histograms  of the classification probabilities computed with the {\it Logistic Regression} model (top) and of the standard CDT order parameter $OP_2$ (bottom), measured for several values of $\kappa_0$ across the $A-C$ phase transition. The red data set ($\kappa_0=4.77$), corresponding to the point closest to the transition, was used for the histogram reweighting procedure. Results are shown for $\bar N_{41}=100\mathrm{k}$. $OP_2$ was rescaled to fit in range $[0,1]$.
}
\label{fig_Histograms_Appendix}
\end{figure}

\section*{Appendix 2}

In the CDT Monte Carlo simulations performed for fixed values of the coupling constants $\tilde \kappa_0,\tilde \Delta$ and $\tilde \kappa_4$ (in practice, we use the data generated closest to the phase transition point), one can sample an arbitrary observable $O({\cal T}_i)$ (e.g., a classification/clustering probability computed with a given ML model or a traditional order parameter) extracted from a representative set of MC configurations (triangulations) ${\cal T}_i$, \mbox{$i=1,\ldots,N_{MC}$}. These configurations are generated according to a probability distribution based on the Boltzmann \mbox{weight $\propto \exp(-S(\tilde \kappa_0,\tilde \Delta,\tilde \kappa_4;{\cal T}_i))$}, where $S$ denotes the Regge action \cite{regge}. Using these MC-generated data, one can estimate the expectation value of the observable measured at the fixed values of the CDT coupling constants by the average 
$$
{\langle O\rangle}_{(\tilde \kappa_0,\tilde \Delta,\tilde \kappa_4)}=\frac{1}{N_{MC}}\sum_{i=1}^{N_{MC}}O({\cal T}_i).
$$
In order to compute the susceptibility (variance) of the observable, one can use
$$
\chi(O)=\langle O^2 \rangle - \langle O \rangle^2.
$$

We now wish to use the same MC data, measured for $\tilde \kappa_0,\tilde \Delta$ and $\tilde \kappa_4$, to estimate the mean value (and susceptibility) of the observable for different, but sufficiently close, values of the coupling constants: \mbox{$\tilde \kappa_0 \to \kappa_0$}, \mbox{$\tilde \Delta \to \Delta$}, \mbox{$\tilde \kappa_4 \to \kappa_4 $}. It can be shown \cite{Ferrenberg:1988yz,Bachtis:2020dmf} that this quantity can be accurately estimated by the weighted average
$$
{\langle O\rangle}_{(\kappa_0,\Delta, \kappa_4)}=\frac{\sum_{i=1}^{N_{MC}} O({\cal T}_i)w_i}{\sum_{i=1}^{N_{MC}}w_i},
$$
where
$$
w_i=\exp(-S( \kappa_0, \Delta, \kappa_4;{\cal T}_i) + S(\tilde \kappa_0,\tilde \Delta,\tilde \kappa_4;{\cal T}_i)),
$$
which is known as the {\it histogram reweighting} method.\footnote{Since in the CDT Monte Carlo simulations the lattice cosmological constant $\kappa_4$ is fine-tuned according to $\kappa_4\to \kappa_4^c(\kappa_0,\Delta,\bar N_{41})$, which is an unknown function, we used a linear interpolation to estimate its values for intermediate values of $\kappa_0$ and $\Delta$ between the MC simulation points.}

The major advantage of histogram reweighting is its simplicity and very low computational cost. For example, for the data sets analyzed in this work, the computation of the reweighted mean values required only a fraction of a second, whereas generating new MC data for shifted values of the CDT coupling constants typically takes several weeks. However, for the reweighting method to work properly, the probability distributions (histograms) of  
the observable measured at the original and shifted values of the coupling constants must sufficiently overlap, which is not always the case.

In particular, histogram reweighting worked remarkably well for the classification/clustering probabilities generated by all ML models. We were therefore able to reconstruct the mean values and susceptibilities of these probabilities over the entire range of coupling constants analyzed for the $A-C$ and $B-C_b$ transitions, obtaining results consistent with direct MC measurements. In contrast, the method failed for the traditional CDT order parameters, even within a narrow range around the original coupling constant values.
The difference originates from the degree of overlap between the supports of the histograms measured at different points in the CDT parameter space. For the ML probabilities, the histograms 
corresponding to the coupling constant value closest to the transition point -- which were used for reweighting -- contain a well-sampled distribution spanning the entire $[0,1]$ range and consequently overlap with the values observed for all other coupling constants where the reweighting procedure was applied. In contrast, the histograms of the CDT order parameters exhibit only partial overlap, which renders the reweighting procedure inefficient; see Fig.~\ref{fig_Histograms_Appendix}.

\bibliography{references}

\end{document}